\documentstyle[12pt]{article}   
\def\be{\begin{equation}}
\def\ee{\end{equation}}
\def\beq{\begin{eqnarray}}
\def\eeq{\end{eqnarray}}
\def\n{\nonumber}
\def\bay{\begin{array}}
\def\eay{\end{array}}

\begin{document}

\begin{titlepage}

\title{A Rotating and Radiating Metric}

\author{
Sanjay M. Wagh\dag \,\,and Pradeep S. Muktibodh\dag\ddag
\\
\\
\dag {\footnotesize Central India Research Institute, Post Box 606,
Laxminagar, Nagpur 440 022, India} \\ 
{\footnotesize E-mail : ciri@bom2.vsnl.net.in} \\
\ddag {\footnotesize Department of Mathematics, Hislop College,
Temple Road, Civil Lines, Nagpur 440 001, India} \\ 
}

\maketitle

\begin{abstract}
A non-static solution of Einstein's field equations of General Relativity
representing the gravitational field of
an axisymmetric radiation flow is obtained using
the Eddington or the Kerr-Schild form for the metric. A solution obtained
here manifestly corresponds to the Kerr metric with its mass-parameter, $m$,
being an arbitrary function of the advanced (retarded) null-time coordinate. 
Then, when  $m$ is constant, the solution reduces to the standard Kerr 
metric expressed in terms of the used null coordinate. And, when the
angular momentum parameter, $a$, constant here, is set to zero, the solution
reduces to the Vaidya metric expressed using the corresponding 
null coordinate.
\end{abstract}

\noindent {\em Keywords:} exact solutions -- radiating metric --
rotation \\

\noindent{\em Running head:}
Rotating and Radiating Metric  --- \\

\end{titlepage}

\section{Introduction}
A non-static generalization of the Schwarzschild metric describing the
gravitational field of a radiating star or that of flowing radiation is
the well-known Vaidya metric [1]. As is equally well-known [2], the Kerr 
metric [3] describes
the exterior field outside of a rotating, axisymmetric black hole. 
In fact, it is [3] the only known stationary, asymptotically flat, vacuum 
metric possessing a central gravitating black hole and inherent spacetime
rotation. Then, the next logical step is that of obtaining a suitable
non-static generalization of the Kerr metric which would correspond to the
exterior gravitational field of a rotating and radiating body. This problem
attracted the attention of several workers in the past [4]. However, to
the best of our knowledge, none of the reported solutions can be considered
entirely satisfactory. For example, it is difficult to study the axisymmetric
collapse of radiation shells using these metrics. 

Therefore, it is the purpose of the present paper to investigate this problem
anew. Apart from the obvious purpose of extending the class of known
solutions to the field equations of General Relativity, aspects related to
other important problems such as those related to the axisymmetric collapse of 
radiation shells and the Cosmic Censorship [5] are the real motivation
behind the present work. However, the relevance of the radiating and rotating
metric presented here for such problems will be the subject of an
independent series of publications [6].

In the present paper, we base our derivation of the radiating and rotating
metric on the Eddington or the Kerr-Schild form [2]. It will be clear from our
procedure of derivation that the solution obtained here manifestly corresponds
to the Kerr metric with its mass-parameter, $m$, being an arbitrary function
of the space-time coordiates. Thus, when $m$ is constant, the solution is
the Kerr metric and, when the angular momentum parameter, $a$, constant
here, is set to zero the solution reduces to the Vaidya metric [1].
[We use $G\;=\;c\;=\;1$ units unless explicitly mentioned otherwise.]

\section{Metric, Its Properties and Field Equations}
We begin our derivation with the
following metric of the Eddington or the Kerr-Schild form
\be g_{\alpha\beta}\;=\;\eta_{\alpha\beta}\;-\;2\ell_{\alpha}\ell_{\beta} \ee
where $\eta_{\alpha\beta}$ is the flat Lorentzian metric in Cartesian
coordinates and $\ell_{\alpha}$ is a null-vector with respect to $\eta_{\alpha
\beta}$. Defining $\ell^{\alpha}\;=\;\eta^{\alpha\beta}\ell_{\beta}$, the
inverse metric $g^{\alpha\beta}$ can be easily shown to be
\be g^{\alpha\beta}\;=\;\eta^{\alpha\beta}\;+\;2\ell^{\alpha}\ell^{\beta} \ee
It then follows that the indices of $\ell_{\alpha}$ can be raised or lowered
using either $\eta_{\alpha\beta}$ or $g_{\alpha\beta}$. Further,
\be {\ell^{\alpha}}_{,\beta}\ell_{\alpha}\;=\;0\;=\;\ell_{\alpha,\beta}
\ell^{\alpha}\ee
where a comma denotes an ordinary derivative. Then, it can be shown that
\be \left\{ {{\alpha}\atop {\beta\;\;\tau}} \right\}\,\ell^{\tau}
\;=\; -\;{\ell^{\alpha}}_{,\tau}\ell^{\tau}
\ell_{\beta}\;-\; \ell^{\alpha}\ell_{\beta,\tau}\ell^{\tau} \ee
which implies that
\be L\;=\; -\;{\ell^{\alpha}}_{;\alpha}\;\equiv
\;-\;{\ell^{\alpha}}_{,\alpha} \ee
\be V^{\alpha}\ell_{\alpha} \;=\;0 \;\;\;\;\;\;\;\;\;\;\;\; V^{\alpha}\;=\;
{\ell^{\alpha}}_{;\beta}\ell^{\beta}\;\equiv\; {\ell^{\alpha}}_{,\beta}
\ell^{\beta} \ee
where $\left\{ {{\alpha} \atop {\beta\;\;\tau}} \right\}$ is the
usual Christoffel symbol of the second kind
and the semicolon denotes the metric preserving covariant derivative.

Moreover, it can be shown [2] that $\sqrt{-\,g}\;=\;1$ quite generally,
where $g$ denotes the determinant of the metric. Therefore,
$$\left\{ {{\alpha} \atop {\beta\;\;\alpha}} \right\}\;=
\;\left[ \log \sqrt{-\,g} \right]_{,\beta}\;=\;0 $$
and the Ricci tensor is
\beq R_{\beta\delta} \;&=&\; \left\{ {{\alpha} \atop {\beta\;\;\alpha}}
\right\}_{,\delta}\;-\;
\left\{ {{\alpha} \atop {\beta\;\;\delta}} \right\}_{,\alpha}\;+\;
\left\{ {{\alpha} \atop {\tau\;\;\delta}} \right\}\,\left\{
{{\tau} \atop {\beta\;\;\alpha}} \right\}\;-\;\left\{
{{\alpha} \atop {\tau\;\;\alpha}} \right\}\,\left\{
{{\tau} \atop {\beta\;\;\delta}} \right\}  \n \\
&\equiv &\;-\;\left\{ {{\alpha} \atop {\beta\;\;\delta}} \right\}_{,\alpha}
\;+\;\left\{ {{\alpha} \atop {\tau\;\;\delta}} \right\}\,\left\{
{{\tau} \atop {\beta\;\;\alpha}} \right\}
\eeq

Now, we assume that
\be V^{\alpha}V_{\alpha}\;=\;0 \ee
Then, it can be shown [2] that $V^{\alpha}$ and $\ell^{\alpha}$ must be
proportional to each other. And, we write
\be V^{\alpha} \;\equiv\; {\ell^{\alpha}}_{,\beta}\ell^{\beta}
\;=\;-\,A( x^{\tau})\ell^{\alpha} \ee
where $A( x^{\tau} )$ is a scalar field. We note that the indices of
$V^{\alpha}$ can be raised or lowered using either $\eta_{\alpha\beta}$
or $g_{\alpha\beta}$.

Further, we use the field equations of general relativity in the following
form
\be R_{\beta\delta}\;=\;\kappa T_{\beta\delta}\;-\;{{\kappa T}\over 2}
g_{\beta\delta}\;\;\;\;\;\;\;\;\;\;\;\kappa\;=\;{{-\,8\pi G}\over {c^4}} \ee
where $G$ is Newton's gravitational constant and $T_{\beta\delta}$ is the
energy-momentum tensor whose trace is denoted by $T$.

Now, using properties (3), (5), (6), (8) and (9) and rearranging terms, we
obtain
\beq R_{\beta\delta} \;=\;&-&\,(\ell_{\beta}\ell_{\delta})_{,\alpha\mu}
\eta^{\alpha\mu}\;-\;[(L+A)\ell_{\beta}]_{,\delta}\;-\;[(L+A)\ell_{\delta}
]_{,\beta} \n \\ &+&\;2 [\,2(A\ell^{\alpha})_{,\alpha}\;-\;A^2\;+\;
{\ell^{\mu}}_{,\nu}{\ell^{\nu}}_{,\mu}\;-\;\eta^{\alpha\mu}{\ell^{\nu}}_{,\mu}
\ell_{\nu,\alpha} ]\,\ell_{\beta}\ell_{\delta} \eeq
Multiplying and contracting eq. (11) with $\ell^{\beta}$, expanding the
terms and using properties (3), (5), (6), (8), (9) and eq. (10), we obtain
\beq \ell^{\beta}R_{\beta\delta}\;&=&\;\kappa\ell^{\beta}T_{\beta\delta}\;-\;
{{\kappa T}\over 2}\,\ell_{\delta} \n \\ &=&\;[\,AL\;+\;A^2\;-\;(L+A)_
{,\alpha}\ell^{\alpha}\;-\;\ell^{\beta}\ell_{\beta,\alpha\mu}\eta^{\alpha\mu}
\,]\,\ell_{\delta} \eeq
But, using the above mentioned properties, it can be easily seen that
\beq [\,2(A\ell^{\alpha})_{,\alpha}\;&-&\;A^2 \;+\;{\ell^{\mu}}_{,\nu}
{\ell^{\nu}}_{,\mu}\;-\;\eta^{\alpha\mu}{\ell^{\nu}}_{,\mu}
\ell_{\nu,\alpha}\,]\,\ell_{\delta} \n \\
&=&\;-\,[\,AL\;+A^2\;-\;(L+A)_{,\alpha}\ell^{\alpha}\;-\;\ell^{\nu}\ell
_{\nu,\alpha\mu}\eta^{\alpha\mu}\,]\ell_{\delta} \n \\
&=&\;-\,\ell^{\alpha}R_{\alpha\delta}\;\;=\;-\,\kappa\ell^{\alpha}
T_{\alpha\delta}\;+\; {{\kappa T}\over 2}\ell_{\delta} \eeq
Then, using eqs. (11) and (13) in eq. (10), we obtain
\beq -\,(\ell_{\beta}\ell_{\delta})_{\alpha\mu}\eta^{\alpha\mu}\;&-&\;
[(L+A)\ell_{\beta}]_{,\delta}\;-\;[(L+A)\ell_{\delta}]_{,\beta} \n \\
&=&\kappa (T_{\beta\delta} \;-\; { T\over 2}\eta_{\beta\delta})\;+\;
2\kappa(\ell^{\alpha}T_{\alpha\delta})\ell_{\beta} \eeq
With a choice of the energy-momentum tensor appropriate to the problem at
hand, we now turn to solving the field equations (14).

\section{Equations for the Radiating Metric}

Now, let the {\em directed flow of radiation} mean a distribution of
electromagnetic energy which any chosen local observer finds flowing in
one and only one direction at that chosen location. Then, the energy-
momentum tensor for such a radiation field, as is well-known, is given by
\be T_{\beta\delta}\;=\;\zeta\,\ell_{\beta}\ell_{\delta} \;\;\;\;
T\;\equiv \;T^{\alpha}_{\alpha}\;=\;0\;\;\;\;\;\zeta\;=\;{{\sigma}\over
{\ell_0^2}} \ee
where $\sigma$ is the density of radiation and the lines of flow of
radiation are the null geodesics of the metric $g_{\alpha\beta}$.
It may be noted that $\ell_o$ is a dimensionless quantity and, hence, 
$\zeta$ and $\sigma$ have the dimensions of energy density, both. However, 
as will be seen later, it is $\sigma$ that is directly related to the
flux of radiation.

Then, for the traceless energy-momentum tensor of eq. (15), we note that
$\ell^{\alpha}T_{\alpha\delta}\;=\;0$. Therefore, the field equations,
eq. (14), become
\be -\;\Box^2(\ell_{\beta}\ell_{\delta})\;-\;[(L+A)\ell_{\beta}]_{,\delta}
\;-\;[(L+A)\ell_{\delta}]_{,\beta}\;=\;{{\kappa\sigma}\over{\ell_o^2}}
\ell_{\beta}\ell_{\delta} \ee
where $\Box^2\;=\;(\partial^2 /\partial x^{o\,2})\;-\;\nabla^2$ is the
standard D'Alembertian operator.

Now, introduce a three-vector $\lambda_i$ as (latin indices to take values
1, 2, 3)
\be \ell_{\alpha}\;=\;\ell_o\,(1,\;\lambda_i),\;\;\;\;\;\;\;\;\;\;
\ell^{\alpha}\;=\;\ell_o\,(1,\;-\,\lambda^i) \ee
such that $\lambda_i$ is a flat-space unit vector, $\lambda^i\lambda_i\;=\;
1$ since $\ell^{\alpha}$ is a flat-space null vector. Expressed in terms
of $\lambda_i$, the field equations are
\begin{eqnarray*}
(\beta\,=\,\delta\,=\,0) \\ &- \,\Box^2(\ell_o^2)\;-\;
2\,[(L+A)\ell_o]_{,o}\;=\;\kappa\sigma  &\hfill(18A) \\ 
(\beta\,=\,0,\;\delta\,=\,i) \\ &- \,\Box^2(\ell_o^2\lambda_i)
\;-\;[(L+A)\ell_o]_{,i}\;-\;[(L+A)\ell_o\lambda_i]_{,o}\;=\;\kappa\sigma
\lambda_i &\hfill(18B) \\ 
(\beta\,=\,i,\;\delta\,=\,j) \\ &- \,\Box^2(\ell_o^2\lambda_i\lambda_j)
\;-\;[(L+A)\ell_o\lambda_i]_{,j}\;-\;[(L+A)\ell_o\lambda_j]_{,i}
\;=\;\kappa\sigma\lambda_i\lambda_j &\hfill(18C) 
\end{eqnarray*}

\setcounter{equation}{18}
Now, using the explicit form of the D'Alembertian and
\be \lambda_{i,o}\;=\;0 \ee
the eq. (18C) can be manipulated with the help of eqs. (18A) and (18B)
to yield
\be \lambda_{i,k}\lambda_{j,k}\;=\;\left( {{L+A}\over {2\ell_o}} \right)
\left[ \lambda_{i,j}\;+\;\lambda_{j,i} \right] \ee
Here and hereafter, we sum over any repeated latin index regardless of
position, for example, index $k$ is to be summed over in eq. (20).

Now, using eq. (5) and eq. (6) with $\alpha\;=\;0$, we get
\be L+A \;=\; \ell_o\lambda_{i,i}\;+\;2\,\ell_{o,i}\lambda_i\;-\;
2\,\ell_{o,o} \ee
The form of the last two terms of eq. (21) then suggests a substitution
\be \ell_o^2 \;=\; m\left( x^{\alpha} \right)\,g \left( x^k \right) \ee
which gives
$$ {{L+A}\over {\ell_o}}\;=\;\lambda_{i,i}\;+\;{{g_{,i}\lambda_i}\over g}
\;+\;{1\over m}\,\left[ m_{,i}\lambda_i\;-\;m_{,o} \right] \eqno(21A) $$
Demanding $(L+A)/\ell_o$ to be independent of $m$, we then require
\be m_{,o}\;=\;m_{,i}\lambda_i \ee
so that
$$ {{L+A}\over {\ell_o}}\;=\;\lambda_{i,i}\;+\;{{g_{,i}\lambda_i}\over g}
\;\equiv\;2\,p\left( x^k \right), \;\;\;{\rm say} \eqno(21B) $$
\setcounter{equation}{23}

Then, eq. (20) is the same as the one which arises in the stationary case,
$m\;=\;$constant and can be solved in the manner as in [2]. Therefore,
without going into the details, which can be found in [2], we note that
the field eq. (20) is equivalent to
\be \nabla^2\gamma\;=\;0,\;\;\;\;\;\;\;\;\left( \nabla\omega \right)^2\;=\;
1 \ee
where
\be \gamma\;=\;\alpha\;+\;\imath \beta,\;\;\;\;\;\;\alpha\;=\;p\,\left( 1 \,-\,
\cos \vartheta \right),\;\;\;\;\;\;\beta\;=\;p\,\sin \vartheta,\;\;\;\;\;\;
\omega\;=\;{1\over {\gamma}} \ee
( Here $\vartheta$ is the angle of rotation which brings $\lambda_i$
along the $X$-axis. ) Depending upon consistent boundary conditions
the above Laplace and the Eikonal equations, eq. (24), can be solved and
the unit vector $\vec{\lambda}$ can be obtained as
\be \vec{\lambda}\;=\;{ {\nabla\omega\;+\;\nabla\omega^{\star}\;-\;
\imath \left( \nabla\omega\,\times\,\nabla\omega^{\star} \right)}\over
{1\;+\;\nabla\omega \bullet \nabla\omega^{\star}} } \ee
which satisfies eq.(20) or equivalently,
\be \lambda_{i,k}\;=\;\alpha\,\left( \delta_{ik}\;-\;\lambda_i\lambda_k
\right) \;+\;\beta\,\epsilon_{ikl}\lambda_l \ee

Moreover, we also note at this point that when $m\;=\;$constant,
\be g\left( x^k \right) \;=\;\alpha \ee
is the solution to field equations, eqs. (18A) - (18C), unique upto
a multiplicative constant [2]. Then, from eq. (24), $g_{,kk}\;\equiv\;\nabla^2
\alpha\;=\;0$, quite generally.

Therefore, having determined $\vec{\lambda}$ and having noted that
$g\left( x^k \right) \;=\;\alpha$, we now turn to eqs. (18A) and (18B).
For this purpose, rewrite eq. (18B) in the form
$$ -\,\Box^2\left( \ell_o^2\lambda_i \right)\;-\;2\,\left( p\,\ell_o^2
\right)_{,i}\;-\;2\,p\,\left( \ell_o^2 \right)_{,o}\lambda_i\;=\;
\kappa\sigma\lambda_i $$
expand the D'Alembertian and use eqs. (18A) and (22) to obtain
\be m \,\left[ g\nabla^2\lambda_i\;+\;2\,\lambda_{i,k}g_{,k} \;-\;2\,
\left( p\,g \right)_{,i} \right] \;+\;2\,p\,g\,\left[ m_{,o}\lambda_i
\;-\;m_{,i} \right] \;+\;2\,g\,\lambda_{i,k}m_{,k}\;=\;0 \ee
Then, we can use eq. (27) to simplify terms involving $\lambda_{i,k}$ and
reduce eq. (29) to
\be 2\,\alpha\,p\,\cos \vartheta \,\left[ m_{,o}\lambda_i\;-\;m_{,i}
\right]\;=\;0 \ee
where we have used $\;\;g\left( x^k \right)\;=\;\alpha\;\;$ and
$\;\;\nabla^2\left( \alpha\lambda_i \right)\;=\;\left( \alpha^2\;
+\;\beta^2 \right)_{,i}\;\;$ as can be easily verified [2].
Now, multiplying and contracting eq. (30) with $\;\lambda_i\;$, we see that
eq. (30) is equivalent to eq. (23).

We now use $\;\;m_{,oo}\;-\;m_{,kk}\;=\;-\,m_{,o}\lambda_{k,k},\;\;$
which follows
from eq. (23), and substitute for $\;\lambda_{k,k}\;$ from eq. (21B) to
rewrite eq. (18A) as
\be \kappa\sigma\;=\;m_{,o}\,\left( \alpha_{,k}\lambda_k\;-\;2\,p\,\alpha
\right) \ee
But, differentiating eq. (27) with respect to $x^k$ and manipulating the
terms gives
$$\alpha_{,k}\lambda_k\;=\;\beta^2\;-\;\alpha^2\;=\;2\,p\,\alpha\,\cos
\vartheta $$
and, hence, eq. (31) reduces to
\be \sigma\;=\;{{\alpha^2}\over{4\,\pi}}\,m_{,o} \;\;\;\;\;\;\;\;\;\;\;\;\;\;
\zeta\;\equiv\;{{\sigma}\over{\ell_o^2}}\;=\;{{\alpha}\over{4\pi}}\,
{{m_{,o}}\over m} \ee
Therefore, we have reduced the field equations, eqs. (18), to eqs. (23),
(24) and (32).

It is important to note that by demanding in eq. (21A) that $(L+A)/\ell_o$ 
to be independent of $m$, we have effectively decoupled eq. (20) from
eqs. (30) and (32). That is to say, eqs. (30) and (32) are not required to
determine the solution of eq. (20) which basically decides the symmetry of
the solution. Further, we see that the equations yield the standard
(Kerr) solution for $m\;=\;$constant [2].

We now turn to a specific solution to the field equations, eqs. (23), (24)
and (32) that has the symmetry of the Kerr metric.

\section{The Radiating and Rotating Metric}

Let us consider eq. (24). As is well-known [2,3], the following
is the solution of eq. (24) corresponding to the Kerr metric
\beq
\gamma\;=\;\left[ x^2\;+\;y^2\;+\;\left( z\,-\,\imath a \right)^2 \right]
^{-\,1/2} \n \\ \n \\
g\left( x^k \right)\;\equiv\;\alpha\;=\;{{\rho^3}\over{\rho^4\,+\,a^2\,z^2}},
\;\;\;\;\;\;\;\rho^2\;=\;{{r^2\,-\,a^2}\over 2}\;+\;\left[ {{\left( r^2\,-\,a^2
\right)^2}\over 4}\;+\;a^2\,z^2 \right]^{1/2} \\ \n \\
\lambda_x\;=\;{{\rho\,x\,+\,a\,y}\over{\rho^2\,+\,a^2}},\;\;\;\;\;\;\;\;
\lambda_y\;=\;{{\rho\,y\,-\,a\,x}\over{\rho^2\,+\,a^2}},\;\;\;\;\;\;\;\;
\lambda_z\;=\;{z\over {\rho}} \n
\eeq
where $\;\;r^2\;=\;x^2\,+y^2\,+z^2\;\;$ with $x$, $y$, $z$ being
the Cartesian coordinates
and $a$, constant here, is the measure of the angular momentum per unit
mass, $m$, of the central gravitating object. We note that in choosing
the form of $\gamma$ above, we have lost no generality [2]. As is
well-known, this solution reduces to the spherically symmetric case
when $a\;=\;0$.

Further, it is easy to see that eq. (23) is equivalent to the existence of
a new time coordinate, say, $v$, which satisfies
\be v_{,o}\;=\;v_{,i}\lambda_i \ee
As is well-known, this is the relation satisfied by the choice $v\;=\;
t\;+\;\rho$.

Now, in terms of the advanced null coordinate, $v$, eq. (32) is
\be \sigma \;=\;{{\alpha^2}\over {4\,\pi}}\,m_{,v} \;\;\;\;\;\;\;
\zeta\;=\;{{\alpha}\over{4\pi}}\,{{m_{,v}}\over m}\;\;\;\;\;\;\;
m_{,v}\;\geq \;0 \ee
where $m\;\equiv\;m(v)$ is an arbitrary, non-negative, increasing function
of $v$. We note that eq. (35) then relates the density of radiation to the
rate of change of mass, $m$, with respect to advanced null time $v$ and can
be taken as a defining relation. Therefore, we have completed our solution
of the field equations, eqs. (18), or, equivalently, that of eqs. (23), (24)
and (32).

Then, putting all the pieces together, the metric can be explicitly displayed
in the $\left( v,\;\rho,\;\theta,\;\phi \right)$ coordinates in the 
Carmeli-Kaye form as
\beq
ds^2\;=\;\left( 1\;-\;{{2\,m\,\rho}\over{\rho^2\;+\;a^2\,\cos^2 \theta}}
\right)\,dv^2\;-\;\left( \rho^2\;+\;a^2\,\cos^2 \theta \right)\,d\theta^2
\;-\;2\,dv\,d\rho\n \\
-\;2\,a\,\sin^2 \theta\,d\rho\,d\phi
\;-\;{{4\,m\,\rho\,a\,\sin^2 \theta}\over{\rho^2\;+\;a^2\,\cos^2 \theta}}\,
dv\,d\phi\n \\
-\;\left[ \rho^2\;+\;a^2\;+\;{{2\,m\,\rho\,a^2\,\sin^2 \theta}
\over{\rho^2\;+\;a^2\,\cos^2 \theta}} \right]\,\sin^2 \theta\,d\phi^2
\eeq
where $m\;\equiv\;m(v)$ and we have changed the coordinates as
\beq
\cos \theta \;=\;{z \over {\rho}}\;\;\;\;\;x\;+\;\imath\,y\;=\;
\left( \rho\;-\;\imath\,a \right)\,e^{\imath\,\phi}\,\sin \theta \n \\
d\tilde{\phi}\;=\;d\phi\;+\;{{2\,a}\over {\rho^2\;+\;a^2}}\,d\rho
\eeq
and dropped the tilde over $\phi$ while displaying the metric. Clearly,
for $m\;=\;$constant, this is the Kerr metric. And, for $m\;\equiv\;m(v)$
but $a\;=\;0$, this reduces to the Vaidya metric. In this last case, 
$\alpha\;=\;{1 \over r}$ and, hence, $\sigma\;=\;m_{,v}/4\pi r^2$ while
$\zeta\;=\;m_{,v}/4\pi r m$. Clearly, it is $\sigma$ that is directly 
related to the flux of radiation. 

We note that since $m_{,v}\;\geq\;0$ the metric of eq. (36) corresponds to
a situation of collapsing radiation shells. Then, the metric corresponding
to expanding radiation shells can be obtained if we choose $\ell_{\alpha}
\;=\;\ell_o\,\left( 1,\, -\,\vec{\lambda} \right)$ in the place of
$\ell_{\alpha}$ as in eq. (17). In this case, we shall then obtain $m_{,u}\;
\leq\;0$ where $u\;=\;t\;-\;\rho$ is the retarded null time. Of course,
the metric of eq. (36) will then have to be written in terms of the new
$\left( u, \,\rho, \,\theta, \,\phi \right)$ coordinates accordingly but
we shall not write that form of the metric explicitly here.

\goodbreak
\section{Concluding Remarks}

In conclusion, we have obtained here a non-static solution of the field
equations of General Relativity representing the gravitational field of
axisymmetric radiation flow beginning with a metric of the Eddington or
the Kerr-Schild form. The solution manifestly possesses the symmetry of 
the Kerr metric.

\end{document}